\documentclass[a4paper,11pt]{article}
\usepackage{jheppub} 

\usepackage[T1]{fontenc} 

\title{\boldmath The Cosmological Impact of Brane-Chern-Simons Massive Gravity}

\author[a, b]{S. Kazempour,\note{Corresponding author.}}
\author[b]{and A. R. Akbarieh}

\affiliation[a]{School of Physics, Beijing Institute of Technology, Beijing, 100081, China}
\affiliation[b]{Faculty of Physics, University of Tabriz, Tabriz 51666-16471, Iran}

\emailAdd{sobhan.kazempour1989@gmail.com}
\emailAdd{am.rezaei@tabrizu.ac.ir}

\abstract{In this paper, we present a novel extension of massive gravity theory; the Brane-Chern-Simons massive gravity theory. We explore the cosmological implications of this theory by deriving the background equations and demonstrating the existence of self-accelerating solutions. Interestingly, our theory suggests the existence of self-accelerating mechanisms that originate from an effective cosmological constant, leading to intriguing possibilities for understanding the nature of cosmic acceleration. Furthermore, we perform a tensor perturbation analysis to investigate the propagation of gravitational waves in this framework. We derive the dispersion relation for gravitational waves and study their behavior in the Friedmann-Lema\^itre-Robertson-Walker cosmology within the context of Brane-Chern-Simons massive gravity.}

\newcommand{\PlanckMass}{M_{\rm Pl}}  
\begin{document} 
\maketitle
\flushbottom

\section{Introduction}\label{sec:intro}

The late-time accelerated expansion of the Universe remains one of the most intriguing puzzles in modern cosmology, prompting extensive research into the underlying mechanisms driving this phenomenon. While the inclusion of cosmological constant or dark energy components within the framework of general relativity provides a plausible explanation \cite{Weinberg:1988cp,Peebles:2002gy,Copeland:2006wr,Carroll:2003st,Cai:2009zp,Bamba:2012cp}, alternative approaches have also gained significant attention.

Modified gravity theories offer a compelling avenue to explore, encompassing a wide range of models that modify the geometric sector of general relativity \cite{Nojiri:2010wj,Clifton:2011jh,CANTATA:2021ktz,Ishak:2018his,Abdalla:2022yfr}. Within this broad category, curvature-based gravity theories, such as $f(R)$ gravity \cite{Starobinsky:1980te}, $f(G)$ gravity \cite{Nojiri:2005jg}, $f(P)$ gravity \cite{Erices:2019mkd}, Lovelock gravity \cite{Lovelock:1971yv}, and Horndeski/Galileon scalar-tensor theories \cite{Horndeski:1974wa,Deffayet:2009wt}, have been extensively studied. Another branch of modified gravity involves torsion-based theories, including $f(T)$ gravity \cite{Bengochea:2008gz,Cai:2015emx}, $f(T, T_{G})$ gravity \cite{Kofinas:2014owa}, $f(T, B)$ gravity \cite{Bahamonde:2015zma}, and scalar-torsion theories \cite{Geng:2011aj}.

In this work, we focus on a particular subclass of gravitational modification known as massive gravity, where the graviton is endowed with a mass \cite{deRham:2010ik,deRham:2010kj,Hinterbichler:2011tt,Hassan:2011hr,Hassan:2011zd}. The concept of massive gravity was first introduced by Fierz and Pauli in 1939, who formulated a unique Lorentz-invariant linear theory without ghosts in a flat spacetime \cite{Fierz:1939ix}. However, subsequent work by van Dam, Veltman, and Zakharov revealed the presence of a discontinuity (known as the van Dam-Veltman-Zakharov discontinuity) when taking the massless limit of the theory \cite{vanDam:1970vg,Zakharov:1970cc}. Vainshtein proposed extending the theory to the nonlinear level to address this issue \cite{Vainshtein:1972sx}, but this led to the discovery of a ghost instability, dubbed the Boulware-Deser ghost \cite{Boulware:1972yco}.

Significant progress was made by de Rham, Gabadadze, and Tolley (dRGT) who developed a fully nonlinear massive gravity theory without the Boulware-Deser ghost in a certain decoupling limit \cite{deRham:2010ik, deRham:2010kj}. While dRGT massive gravity successfully explains the accelerated expansion of the Universe in an open Friedmann-Lema\^itre-Robertson-Walker (FLRW) geometry, it faces challenges in accommodating a homogeneous and isotropic Universe \cite{DeFelice:2012mx}. Additionally, the theory suffers from a strong coupling problem and a nonlinear ghost instability, causing scalar and vector perturbations to vanish \cite{Gumrukcuoglu:2011zh}.

To address these issues, the quasi-dilaton massive gravity theory was introduced \cite{DAmico:2012hia, Gannouji:2013rwa}, with further developments in subsequent works \cite{DeFelice:2013dua,Mukohyama:2014rca,Tamanini:2013xia,EmirGumrukcuoglu:2014uog,Cai:2013lqa, Kahniashvili:2014wua,Cai:2014upa,Gumrukcuoglu:2016hic,Gumrukcuoglu:2017ioy,Gumrukcuoglu:2020utx, DeFelice:2021trp,Akbarieh:2021vhv,Akbarieh:2022ovn,Kazempour:2022let,Kazempour:2022giy,Kazempour:2022xzy}. However, even in this extended framework, instabilities persist in the perturbation analysis \cite{Gumrukcuoglu:2013nza,Haghani:2013eya,DAmico:2013saf}.

In this paper, we propose a novel extension of massive gravity, dubbed Brane-Chern-Simons massive gravity, which exhibits a self-accelerating solution and instability-free perturbation. Brane-Chern-Simons gravity is a theoretical framework that synergizes brane theory and Chern-Simons theory, offering a unique perspective on the nature of gravity and its interplay with other forces \cite{Izaurieta:2009hz,Gomez:2020agz}. By combining the concepts of branes in higher-dimensional spaces and the dynamics of gauge fields in odd-dimensional spaces, Brane-Chern-Simons gravity suggests a topological origin of gravity and provides a rich framework for exploring new physics beyond the standard model \cite{Izaurieta:2009hz,Concha:2016kdz,Gomez:2011zzd,Gomez:2020agz}. The other research related to the Chern-Simons Gravity can be found in these references \cite{Nojiri:2019nar,Odintsov:2022hxu,Nojiri:2020pqr,Concha:2014vka}.

The main objectives of this work are to demonstrate the existence of a self-accelerating solution in Brane-Chern-Simons massive gravity which is related to the effective cosmological constant and to perform a perturbation analysis. In particular, we will derive the modified dispersion relation for gravitational waves within this theory.

The paper is structured as follows: In Section \ref{sec:2}, we introduce the Brane-Chern-Simons massive gravity framework, derive the background equations of motion, and extract self-accelerating solutions. Section \ref{sec:3}, delves into the cosmological perturbation analysis, focusing on tensor perturbations and the dispersion relation of gravitational waves. Finally, in Section \ref{sec:5}, we summarize our key findings and discuss future research directions.

\section{\label{sec:2}Brane-Chern-Simons Massive Gravity}

In this section, we introduce the Brane-Chern-Simons massive gravity action, and we discuss the evolution of a cosmological background. The action includes Planck mass $\PlanckMass$,
the Ricci scalar $R$, the constant $\varepsilon$, the potential $V(\varphi)$, the gravitational constant $\kappa$, a dynamical metric $g_{\mu\nu}$ and its determinant $\sqrt{-g}$. The action is given by\\
\begin{eqnarray}\label{Action}
S =\frac{\PlanckMass^{2}}{2}\int d^{4} x \Bigg\{\sqrt{-g}\bigg[R + \varepsilon R V(\varphi) - 2 \kappa V (\varphi) + 2{m}_{g}^{2}U(\mathcal K)\bigg]\Bigg\} + \int d^{4}x \sqrt{-g} \mathcal{L}_{m} ,\nonumber\\
\end{eqnarray}
also, $m_{g}$ is the mass of graviton and $\mathcal{L}_{m}$ is the matter Lagrangian.
In the following, we introduce the $U({\cal K})$ of this action separately.
It is clear that the mass of the graviton comes up with the potential $U$ which consists of three parts.
\begin{equation}\label{Upotential1}
U(\mathcal{K})=U_{2}+\alpha_{3}U_{3}+\alpha_{4}U_{4},
\end{equation}
where $\alpha_3$ and $\alpha_4$ are dimensionless free parameters of the
theory. $U_i$ ($i=2,3,4$) is given by,
\begin{eqnarray}\label{Upotential2}
U_{2}&=&\frac{1}{2}\big([\mathcal{K}]^{2}-[\mathcal{K}^{2}]\big),
\nonumber\\
U_{3}&=&\frac{1}{6}\big([\mathcal{K}]^{3}-3[\mathcal{K}][\mathcal{K}^{2}]+2[\mathcal{K}^{3}]\big),
\nonumber\\
U_{4}&=&\frac{1}{24}\big([\mathcal{K}]^{4}-6[\mathcal{K}]^{2}[\mathcal{K}^{2}]+8[\mathcal{K}][\mathcal{K}^{3}]+3[\mathcal{K}^{2}]^2-6[\mathcal{K}^{4}]\big),
\end{eqnarray}
where the quantity "$[\cdot]$'' is interpreted as the trace of the tensor
inside brackets. It is essential to mention that the building block tensor
$\mathcal{K}$ is defined as
\begin{equation}\label{K}
\mathcal{K}^{\mu}_{\nu} = \delta^{\mu}_{\nu} -
\sqrt{g^{\mu\alpha}\partial_{\alpha}\phi^{c}\partial_{\nu}\phi^{d}\eta_{cd}},
\end{equation}
where $g^{\mu\nu} $ is the physical metric, $\eta_{cd}$ is the Minkowski
metric with $c,d= 0,1,2,3$ and $\phi^{c}$ are the Stueckelberg fields which
are introduced to restore general covariance.
According to our cosmological application purpose, we adopt the general form of Friedman-Lema\^itre-Robertson-Walker (FLRW) Universe. 
So, the general expression of the corresponding dynamical metric is given as
follows,
\begin{align}
\label{DMetric}
g_{\mu\nu} d x^{\mu} d x^{\nu} & = -N^{2} d t^{2} + a^{2} \Omega_{i,j} d x^{i} d x^{j}, \\
\label{FMetric}
\Omega_{i,j} d x^{i} d x^{j} & = d x^{2} + d y^{2} + d z^{2} - \frac{|K| (x dx + y dy + z dz)^{2}}{1 + |K| (x^{2} + y^{2} + z^{2})}.
\end{align}
Here it is worth pointing out that $N$ is the lapse function of the
dynamical metric, and it is similar to a gauge function. Also, it is clear
that the scale factor is represented by $a$, and $\dot{a}$ is the
derivative with respect to time. Meanwhile, it should be mentioned that $x^{0}=t$, $x^{1}= x$, $x^{2}=y$, $x^{3}= z$; $\mu, \nu = 0,...,3$; and $i,j= 1,2,3$. Furthermore, $|K|$ is the Gaussian curvature of the space and, the lapse function relates the coordinate-time $dt$ and the proper-time $d\tau$ via $d\tau=Ndt$
\cite{Scheel:1994yr,Christodoulakis:2013xha}. 

Concerning the scalar fields $\phi^{c}, (c= 0, ...,3)$, the following ansatz are considered,
\begin{eqnarray}
\phi^{0} && = f(t) \sqrt{1+ |K| (x^{2} + y^{2} + z^{2})}, \nonumber\\
\phi^{1} && = \sqrt{|K|} f(t) x, \nonumber\\
\phi^{2} && = \sqrt{|K|} f(t) y,\nonumber\\
\phi^{3} && = \sqrt{|K|} f(t) z.
\end{eqnarray}
where function $f(t)$ is the Stueckelberg scalar function whereas $\frac{\partial f(t)}{\partial t}=\dot{f}(t)$ \cite{Arkani-Hamed:2002bjr}.

Therefore, the point-like Lagrangian of the Brane-Chern-Simons massive gravity is
\begin{eqnarray}
\mathcal{L} = -\frac{a \PlanckMass^{2}}{N} \Bigg\lbrace \bigg( 3 |K| + \big( 3 |K| \varepsilon + \kappa a^{2}\big) V(\varphi) \bigg) N^{2} + 3 \big( 1 + \varepsilon V(\varphi) \big) \dot{a}^{2} \Bigg\rbrace + m_{g}^{2} \big( a \nonumber\\ - \sqrt{|K|} f(t) \big) \PlanckMass^{2} \Bigg\lbrace |K| f(t)^{2} \bigg( (\alpha_{3} + \alpha_{4}) N - \alpha_{4} \dot{f}(t) \bigg) + a^{2} \bigg( ( 6 + 4 \alpha_{3} + \alpha_{4} ) N \nonumber\\  - ( 3 + 3 \alpha_{3} + \alpha_{4} ) \dot{f}(t) \bigg) + \sqrt{|K|} a f(t) \bigg(  ( 3 \alpha_{3} + 2 \alpha_{4}) \dot{f}(t) - ( 3 + 5 \alpha_{3} + 2 \alpha_{4} ) N \bigg) \Bigg\rbrace + \mathcal{L}_{m}. \nonumber\\
\end{eqnarray}
We can assume the matter sector $\mathcal{L}_{m}$ to consist of a perfect fluid with the energy density $\rho_{m}$ and pressure $p_{m}$. In order to simplify expressions later, we define
\begin{equation}
H\equiv\frac{\dot{a}}{Na}.
\end{equation}

\subsection{Background Equations of Motion}\label{subsec4}

Without loss of generality, we can assume that $\dot{f}\geq 0$, $f \geq 0$, $a > 0$, and $N > 0$, at least within the vicinity of the time of interest. Consequently, the equation obtained by varying with respect to $f(t)$ encapsulates all the nontrivial information.
In this procedure, a constraint equation can be derived by varying with respect to $f$. So, the equation is given by
\begin{eqnarray}\label{Cons1}
\frac{\delta \mathcal{L}}{\delta f}=  3 m_{g}^{2} \PlanckMass^{2} \big( \dot{a} - \sqrt{|K|}N \big) \bigg\lbrace a^{2} \big( 3 + 3\alpha_{3} + \alpha_{4} \big) - 2 \sqrt{|K|} a f \big( 1 + 2\alpha_{3} +\alpha_{4} \big) \nonumber\\ + |K| f^{2} \big( \alpha_{3} + \alpha_{4} \big) \bigg\rbrace = 0.
\end{eqnarray}
In this stage, the Friedman equation is achieved by varying with respect to
the lapse $N$,
\begin{eqnarray}\label{EqN}
\frac{1}{\PlanckMass^{2} a^{3}}\frac{\delta \mathcal{L}}{\delta N}= 3 H^{2} ( 1 + \varepsilon V(\varphi) ) - \frac{3 |K| ( 1 + \varepsilon V(\varphi) )}{a^{2}} - \kappa V(\varphi) + \frac{m_{g}^{2}}{a^{3}} \bigg[ a^{3} \big( \alpha_{4} + 4 \alpha_{3} +6 \big) \nonumber\\ - 3 a^{2} \sqrt{|K|} f \big( \alpha_{4} + 3 \alpha_{3} + 3 \big) + 3 a f^{2} |K| \big( \alpha_{4} + 2 \alpha_{3} + 1 \big)  - |K|^{\frac{3}{2}} f^{3} \big( \alpha_{4} + \alpha_{3} \big)  \bigg] = \rho_{m}. \nonumber\\
\end{eqnarray}

The equation of motion for $a$ is
\begin{eqnarray}\label{Eqa}
\frac{1}{3\PlanckMass^{2} a^{2}N}\frac{\delta \mathcal{L}}{\delta a} = 3 H^{2} \big[ 1 + \varepsilon V(\varphi) \big] - \kappa V(\varphi) + \frac{2 \dot{H}(1+\varepsilon V(\varphi))}{N}  - \frac{|K|}{r^{2}N^{2}}  \big( 1 + \varepsilon V(\varphi) \big) \nonumber\\ + \frac{m_{g}^{2}}{r^{2}N^{3}}  \Bigg\lbrace 2 \sqrt{|K|} r f N \bigg( 1 + 2 \alpha_{3} + \alpha_{4} - \big( 3 + 3 \alpha_{3} + \alpha_{4} \big) N \bigg) + |K| f^{2} \bigg(  \big( 1 + 2\alpha_{3} \nonumber\\ + \alpha_{4} \big) N  - \alpha_{3} - \alpha_{4} \bigg) + r^{2}N^{2} \bigg( N \big( \alpha_{4} + 4 \alpha_{3} +6 \big) - \big( \alpha_{4} + 3 \alpha_{3} + 3 \big) \bigg) \Bigg\rbrace = p_{m}, \nonumber\\
\end{eqnarray}

where
\begin{eqnarray}
r\equiv\frac{a}{N}.
\end{eqnarray}
In the last part of this subsection, it should be noted that the Stuckelberg field $f$ introduces time reparametrization invariance. So, there is a Bianchi identity which relates the four equations of motion,
\begin{eqnarray}
\frac{\delta S}{\delta \varphi}\dot{\varphi}+\frac{\delta S}{\delta f}\dot{f}-N\frac{d}{dt}\frac{\delta S}{\delta N}+\dot{a}\frac{\delta S}{\delta a}=0.
\end{eqnarray}
\subsection{Self-Accelerating Background Solutions}\label{subsec5}
In this step, we want to discuss solutions. It could be started with the Stueckelberg constraint in Eq. (\ref{Cons1}), so, we have
\begin{eqnarray}\label{Self0}
 \big( \dot{a} - \sqrt{|K|}N \big) \bigg\lbrace a^{2} \big( 3 + 3\alpha_{3} + \alpha_{4} \big) - 2 \sqrt{|K|} a f \big( 1 + 2\alpha_{3} +\alpha_{4} \big) + |K| f^{2} \big( \alpha_{3} + \alpha_{4} \big) \bigg\rbrace = 0, \nonumber\\
\end{eqnarray}
Notice that the above equation has three solutions. The first solution, $\dot{a}= |K|N$, suggests that the physical metric $g_{\mu\nu}$ represents Minkowski spacetime in the open FRW chart. As such, it does not accurately depict the characteristics of our Universe and is thus not a viable solution for describing the cosmos we inhabit \cite{Gumrukcuoglu:2011ew}.
Thus, we eliminate this particular solution, yielding the following expression:
\begin{eqnarray}\label{Self}
 a^{2} \big( 3 + 3\alpha_{3} + \alpha_{4} \big) - 2 \sqrt{|K|} a f \big( 1 + 2\alpha_{3} +\alpha_{4} \big)  + |K| f^{2} \big( \alpha_{3} + \alpha_{4} \big) = 0,
\end{eqnarray}

As a result, the two solutions of Eq. (\ref{Self}) are
\begin{equation}\label{XSA}
f = \frac{a}{\sqrt{|K|}}\varpi_{\pm} , \qquad \varpi_{\pm} = \frac{(1 + 2 \alpha_{3} + \alpha_{4}) \pm \sqrt{1 + \alpha_{3} + \alpha_{3}^{2}- \alpha_{4}}}{(\alpha_{3} + \alpha_{4})}
\end{equation}
It is important to note that these two solutions are invalid when $K$ is set to zero. This aligns with the understanding that there exists no non-trivial flat FRW solution under such conditions.

The Friedman equation (\ref{EqN}) could be written in a different form,
\begin{eqnarray}\label{EqFr1}
3 H^{2} \bigg( 1 + \varepsilon V(\varphi) \bigg) - \frac{3 |K| \big( 1 + \varepsilon V(\varphi) \big)}{a^{2}} = \rho_{g} + \rho_{m} + \kappa V(\varphi).
\end{eqnarray}
It is worth mentioning that the effective energy density from the graviton mass term is
\begin{align}
\rho_{g} = - \frac{m_{g}^{2}}{a^{3}} \bigg[ a^{3} \big( \alpha_{4} + 4 \alpha_{3} +6 \big) - 3 a^{2} \sqrt{|K|} f \big( \alpha_{4} + 3 \alpha_{3} + 3 \big) + 3 a f^{2} |K| \big( \alpha_{4} \nonumber\\ + 2 \alpha_{3} + 1 \big) - |K|^{\frac{3}{2}} f^{3} \big( \alpha_{4} + \alpha_{3} \big)  \bigg].
\end{align}
According to Eq. (\ref{XSA}), the above equation can be written 
\begin{eqnarray}
\rho_{g} = \frac{m_{g}^{2}}{\big( \alpha_{3} + \alpha_{4} \big)^{2}} \bigg\lbrace - (1 + \alpha_{3}) \big( 2 + \alpha_{3} + 2 \alpha_{3}^{2} -3 \alpha_{4} \big) \pm \frac{2 \big( |K| ( 1 + \alpha_{3} + \alpha_{3}^{2} - \alpha_{4}) a^{2}  \big)^{\frac{3}{2}}}{|K|^{\frac{3}{2}} a^{3}} \bigg\rbrace. \nonumber\\
\end{eqnarray}
By employing Eq. (\ref{EqFr1}), we can express the dynamical equation Eq.~(\ref{Eqa}) in an alternative form as follows:
\begin{eqnarray}\label{rS}
- \frac{2 |K| \big( 1 + \varepsilon V(\varphi) \big)}{a^{2}} - \frac{2 \dot{H}(1+\varepsilon V(\varphi))}{N}  = (\rho_{m} + p_{m} ) + (\rho_{g} + p_{g}),
\end{eqnarray}
\begin{eqnarray}
p_{g} = \frac{m_{g}^{2}}{r^{2}N^{3}}  \Bigg\lbrace 2 \sqrt{|K|} r f N \bigg( 1 + 2 \alpha_{3} + \alpha_{4} - \big( 3 + 3 \alpha_{3} + \alpha_{4} \big) N \bigg) + |K| f^{2} \bigg(  \big( 1 + 2\alpha_{3} \nonumber\\ + \alpha_{4} \big) N  - \alpha_{3} - \alpha_{4} \bigg) + r^{2}N^{2} \bigg( N \big( \alpha_{4} + 4 \alpha_{3} +6 \big) - \big( \alpha_{4} + 3 \alpha_{3} + 3 \big) \bigg) \Bigg\rbrace.
\end{eqnarray}
where $p_{g}$ represents the effective pressure contribution of the graviton mass terms.
The equation (\ref{rS}) does not introduce any new insights or information, as it is a direct consequence of the Bianchi identities and the conservation of matter.

Also, we can represent the effective pressure term and the effective density term as shown below:
\begin{eqnarray}
p_{eff} = \frac{m_{g}^{2}(1 + \alpha_{3}) \big( 2 + \alpha_{3} + 2\alpha_{3}^{2} - 3 \alpha_{4} \big)}{( \alpha_{3} + \alpha_{4} )^{2}} \pm \bigg[2 m_{g}^{2} \big( |K| (1 + \alpha_{3} + \alpha_{3}^{2} \nonumber\\ - \alpha^{4}) a^{2}  \big)^{3/2} |K|^{-3/2} (\alpha_{3} + \alpha_{4})^{-2} \mp a \big( |K| + |K| \varepsilon V(\varphi) \big)\bigg] \times a^{-3} - \kappa V(\varphi) + p_{m},
\end{eqnarray}
\begin{eqnarray}
\rho_{eff} =  \frac{m_{g}^{2}}{(\alpha_{3} + \alpha_{4})^{2}} \bigg[ - (1 + \alpha_{3}) \big( 2 + \alpha_{3} + 2 \alpha_{3}^{2} - 3 \alpha_{4} \big) \pm 2 \big( |K| ( 1 + \alpha_{3} + \alpha_{3}^{2} \nonumber\\ - \alpha_{4} ) a^{2} \big)^{3/2} |K|^{-3/2} a^{-3} \bigg] + \rho_{m},
\end{eqnarray}
\begin{eqnarray}
By \quad considering \quad V(\varphi)= - \frac{|K|}{|K|\varepsilon + 2 \kappa a^{2}} \Longrightarrow \qquad p_{eff} = - \rho_{eff}.
\end{eqnarray}

By considering the solutions of the constraint equation (\ref{XSA}) and equation (\ref{rS}), the Friedman equation (\ref{EqFr1}) could be rewritten as,
\begin{eqnarray}\label{EqFr}
3 H^{2} \bigg( 1 + \varepsilon V(\varphi) \bigg) - \frac{3 |K| \big( 1 + \varepsilon V(\varphi) \big)}{a^{2}} = \tilde{\rho}_{g} + \kappa V(\varphi) + \rho_{m},
\end{eqnarray}
where
\begin{eqnarray}\label{2.27}
\tilde{\rho}_{g} = m_{g}^{2}\Lambda_{\pm},
\end{eqnarray}
\begin{eqnarray}\label{2.28}
\Lambda_{\pm} \equiv - \frac{1}{(\alpha_{3} + \alpha_{4} )^{2} } \bigg\lbrace 1 + \alpha_{3} \pm \sqrt{1 + \alpha_{3} + \alpha_{3}^{2} - \alpha_{4}} \bigg\rbrace  \times \bigg\lbrace 1 + \alpha_{3}^{2} - 2 \alpha_{4} \nonumber\\ \pm (1 + \alpha_{3}) \sqrt{1 +\alpha_{3} + \alpha_{3}^{2} -\alpha_{4}} \bigg\rbrace.
\end{eqnarray}

Equation (\ref{EqFr}) is analogous to the Friedmann equation describing an open Universe influenced by arbitrary matter and an effective cosmological constant $\tilde{\rho}_{g} = m_{g}^{2}\Lambda_{\pm}$. This cosmological constant arises from the massive graviton term, contributing to the dynamics of the Universe's expansion.

A key finding of this subsection is the identification of self-accelerating solutions within the theory, characterized by an effective cosmological constant, $m_{g}^{2}\Lambda_{\pm}$. Importantly, these solutions do not exhibit strong coupling issues, and they provide a well-behaved description of the accelerated expansion of the Universe. This suggests that the Brane-Chern-Simons massive gravity theory offers a viable framework for explaining the late-time cosmic acceleration without encountering the challenges typically associated with strong coupling regimes. In the absence of the massive gravity component, i.e., when $m_{g}=0$, the Friedmann equation reduces to the equation of state of standard general relativity, incorporating a scalar potential term, and a possible curvature term. Furthermore, when $m_{g}=0$, $\varepsilon =0$ and $\kappa=0$, the theory recovers the standard general relativity framework.
On the background level, our model is equivalent to the standard Lambda-Cold Dark Matter (LCDM) model, which is known to provide an adequate fit to the observational data. Notably, a similar analysis of LCDM model parameter constraints using the Supernova Ia dataset has been previously reported by the Supernova Cosmology Project \cite{SupernovaCosmologyProject:2008ojh}.

\section{Perturbation Analysis}\label{sec:3}

In this section, we would like to analyze tensor perturbation in order to calculate the mass of graviton for our theory which we introduced in the previous section. Furthermore, we are trying to show the stability condition of the system.
In order to find the action for quadratic perturbation, the physical metric
is expanded in small fluctuation, $\delta g_{\mu\nu}$, around a solution
$g_{\mu\nu}^{(0)}$,
\begin{equation}
g_{\mu\nu}=g_{\mu\nu}^{(0)}+\delta g_{\mu\nu}.
\end{equation}
In the following analysis, we keep terms to quadratic order in $\delta
g_{\mu\nu}$. As we demonstrate all analysis in the unitary gauge, there are
not any problems concerning the form of gauge invariant combinations.
Moreover, we write the actions expanded in the Fourier domain with plane
waves, i.e., $\vec{\nabla}^{2}\rightarrow -k^{2}$, $d^{3}x\rightarrow
d^{3}k$. We raise and lower the spatial indices on perturbations by
$\delta^{ij}$ and $\delta_{ij}$.
We start by considering tensor perturbations around the background,
\begin{equation}
\delta g_{ij}=a^{2}h_{ij}^{\rm TT},
\end{equation}
where
\begin{equation}
\partial^{i}h_{ij}=0 \quad {\rm and} \quad g^{ij}h_{ij}=0.
\end{equation}
The tensor perturbed action in the second order can be calculated for each part of the action separately.
We write the Brane-Chern-Simons part of the perturbed action in quadratic order
\begin{eqnarray}
S^{(2)}_{\rm B-C-S}=\frac{\PlanckMass^{2}}{8}\int d^{3}k \, dt \, a^{3} N \Bigg\lbrace \big[ 1 + \varepsilon V(\varphi) \big] \bigg[\frac{\dot{h}_{ij}\dot{h}^{ij}}{N^{2}} - |K| (\frac{k^{2}}{a^{2}}+\frac{4\dot{H}}{N}+6H^{2})h^{ij}h_{ij}\bigg] \nonumber\\ - \frac{2}{\PlanckMass^{2}} \mathcal{L}_{m} + 2 \kappa V(\varphi) h^{ij}h_{ij}\Bigg\rbrace . \nonumber\\
\end{eqnarray}
The second-order piece of the massive gravity sector of the perturbed action can be written as
\begin{eqnarray}
S^{(2)}_{\rm massive}= -\frac{\PlanckMass^{2}}{4}\int d^{3}k \, dt \, m_{g}^{2} ( a - \sqrt{|K|} f )\Bigg\lbrace |K| f^{2} \bigg[ \big( \alpha_{3} + \alpha_{4} \big) N - \alpha_{4} \dot{f} \bigg] + a^{2} \bigg[ \big( 6 + 4 \alpha_{3} \nonumber\\ + \alpha_{4} \big) N - \big( 3 + 3 \alpha_{3} + \alpha_{4} \big) \dot{f} \bigg] + \sqrt{|K|} a f \bigg[ \big( 3 \alpha_{3} + 2 \alpha_{4} \big) \dot{f} - \big( 3 + 5\alpha_{3} + 2 \alpha_{4} \big) N \bigg]  \Bigg\rbrace h^{ij}h_{ij}. \nonumber\\
\end{eqnarray}
Summing up the second order pieces of the perturbed actions $S^{(2)}_{\rm B-C-S}$, and $S^{(2)}_{\rm massive}$ we obtain the total action in a second order for tensor perturbations
\begin{eqnarray}
S^{(2)}_{\rm total}=\frac{\PlanckMass^{2}}{8}\int d^{3}k \, dt \, a^{3} N \bigg\lbrace \big[ 1+\varepsilon V(\varphi) \big] \frac{\dot{h}^{ij}\dot{h}_{ij}}{N^{2}} - \Big( \big[ 1+\varepsilon V(\varphi) \big] |K| \frac{k^{2}}{a^{2}}+M_{\rm GW}^{2}\Big)h^{ij}h_{ij}\bigg\rbrace .\nonumber\\
\end{eqnarray}
So, the dispersion relation of gravitational wave is
\begin{eqnarray}
M^{2}_{\rm GW}= \big( 4\frac{\dot{H}}{N} + 6 H^{2} \big) \big[ 1+\varepsilon V(\varphi) \big] |K| + \frac{2}{\PlanckMass^{2}}p_{m} - 2 \kappa V(\varphi) + \Upsilon , \nonumber\\
\label{eq:M2:GW1}
\end{eqnarray}
where

\begin{eqnarray}\label{upsilon}
\Upsilon = && \frac{m_{g}^{2}}{|K|^{4} (\alpha_{3} + \alpha_{4})^{4}\sqrt{|K| \zeta a}} \times \bigg[ - |K| ( \alpha_{3} + \alpha_{4}) + \sqrt{|K|} ( 1 + 2\alpha_{3} + \alpha_{4} ) a \pm \sqrt{|K| \zeta a} \bigg] \Bigg\lbrace |K|^{3} ( \alpha_{3} \nonumber\\ && + \alpha_{4} )^{3} (6 + 4 \alpha_{3} + \alpha_{4}) \sqrt{|K| \zeta a} \mp 2 H |K|^{2} (\alpha_{3} + \alpha_{4})^{2} ( 3 + 3 \alpha_{3} + \alpha_{4}) a^{2} \bigg( \sqrt{|K|} (1 + \alpha_{3} \nonumber\\ && + \alpha_{3}^{2} - \alpha_{4}) a  \pm (1 + 2 \alpha_{3} + \alpha_{4} ) \sqrt{|K| \zeta a} \bigg) + \bigg( \pm \sqrt{|K|} ( 1 + 2\alpha_{3} + \alpha_{4}) a + \sqrt{|K| \zeta a} \bigg)^{2}  \nonumber\\ && \times  \bigg(  |K|  (\alpha_{3} + \alpha_{4})^{2} \sqrt{|K| \zeta a } -  2 H \alpha_{4} a^{2} \big( \pm \sqrt{|K|} \zeta + ( 1 + 2 \alpha_{3} + \alpha_{4} ) \sqrt{|K|\zeta a}  \big)  \bigg) \nonumber\\ && - |K| ( \alpha_{3} + \alpha_{4}) \bigg( \sqrt{|K|} (1 + 2\alpha_{3} + \alpha_{4}) a \pm \sqrt{|K| \zeta a} \bigg) \times \bigg(  |K| ( \alpha_{3} + \alpha_{4}) (3 \nonumber\\ && + 5 \alpha_{3} + 2\alpha_{4}) \sqrt{|K| \zeta a}  \mp  2 H (3 \alpha_{3} + 2\alpha_{4}) a^{2} \big[ \sqrt{|K|} \zeta \nonumber\\ && \pm (1 + 2\alpha_{3} + \alpha_{4}) \sqrt{|K| \zeta a}  \big] \bigg)  \Bigg\rbrace , \qquad where \qquad \zeta =  a (1 + \alpha_{3}  +\alpha_{3}^{2} - \alpha_{4}) .
\end{eqnarray}

\begin{figure}
\centering
\includegraphics[width=13cm]{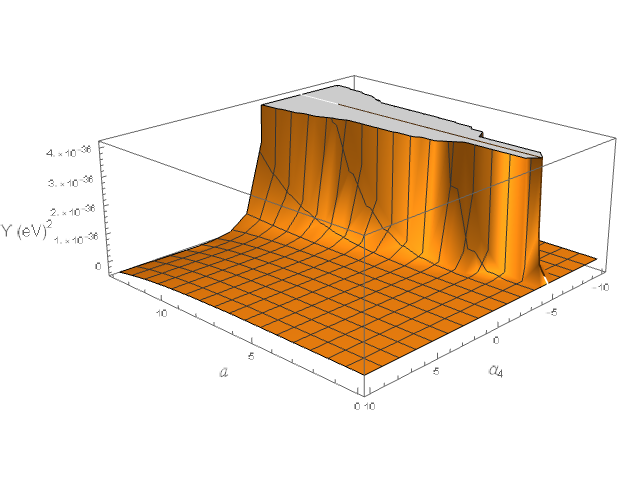}
\caption[The possible ranges of $\gamma$, scale factor $a$, and $\alpha_{4}$ are determined by considering the values for $|K|=1$, $m_{g}= 10^{-23} eV$,  $H=69 (km/s)/Mpc$, and $\alpha_{4} = 5$.]
{The possible ranges of $\gamma$, scale factor $a$, and $\alpha_{4}$ are determined by considering the values for $|K|=1$, $m_{g}= 10^{-23} eV$ \cite{LIGOScientific:2017bnn},  $H=69 (km/s)/Mpc$, and $\alpha_{3} = 5$.}\label{fig9}
\end{figure}

\begin{figure}
\centering
\includegraphics[width=13cm]{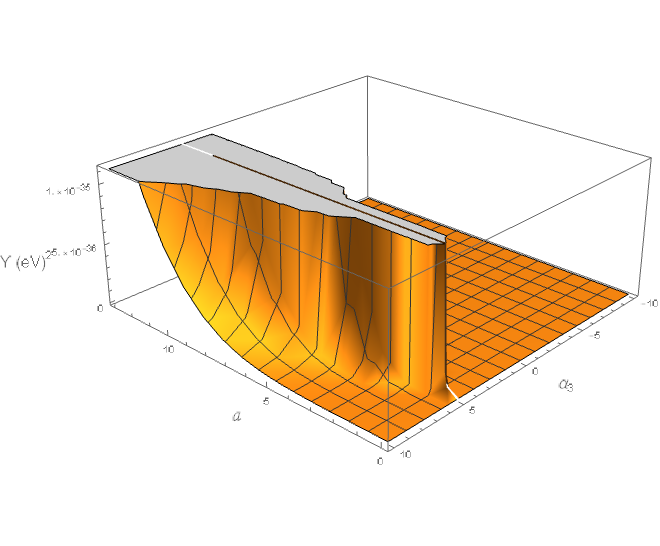}
\caption[The possible ranges of $\gamma$, scale factor $a$, and $\alpha_{3}$ are determined by considering the values for $|K|=1$, $m_{g}= 10^{-23} eV$,  $H=69 (km/s)/Mpc$, and $\alpha_{4} = -5$.]
{The possible ranges of $\gamma$, scale factor $a$, and $\alpha_{3}$ are determined by considering the values for $|K|=1$, $m_{g}= 10^{-23} eV$,  $H=69 (km/s)/Mpc$, and $\alpha_{4} = -5$.}\label{fig10}
\end{figure}

The equation (\ref{upsilon}) is derived from substituting the solutions of $f$ from the equation (\ref{XSA}). In Figures (\ref{fig9}) and (\ref{fig10}), we show the possible ranges of Eq. (\ref{upsilon}) which is the dispersion relation of gravitational wave originating from the massive gravity part. 
The sign of the mass square of gravitational waves plays a crucial role in ensuring the stability of long-wavelength gravitational waves. When the mass square is positive, it indicates a stable configuration. Conversely, a negative mass square implies the presence of tachyons, leading to potential instability. However, due to the characteristic timescale associated with the tachyonic mass, which is on the order of the Hubble scale, the development of any instability would take a duration comparable to the age of the Universe, thus having minimal impact on the stability of the system over observable timescales.

The calculation of the dispersion relation for gravitational waves in the Brane-Chern-Simons massive gravity theory provides a representation of the propagation of gravitational perturbations in the Friedmann-Lemaître-Robertson-Walker (FLRW) cosmology. In principle, this propagation can be tested through observations of cosmological events, particularly those involving gravitational waves. The modifications introduced by this theory will result in additional contributions to the phase evolution of gravitational waveforms, which can be detected using precise matched-filtering techniques in data analysis \cite{Will:1997bb, Mirshekari:2011yq}.

Following the groundbreaking discovery of gravitational waves from a merging binary black hole (GW150914) by the LIGO/Virgo Collaboration, interest in testing the mass of the graviton has been rekindled \cite{LIGOScientific:2016lio, LIGOScientific:2019fpa,LIGOScientific:2020tif,Shao:2020shv}. Recent constraints on the graviton mass, derived from the combined analysis of gravitational wave events from the first and second transient catalogs, yield an upper limit of $m_{g} \leq 1.76 \times 10^{-23} \, \mathrm{eV} / c^{2}$ at 90\% credibility \cite{LIGOScientific:2020tif}. Although the corresponding Compton wavelength is still much smaller than the Hubble scale, limiting its relevance to modified cosmology, ongoing and future efforts will continue to test this fundamental aspect of gravitation with an increasing number of gravitational wave events at various wavelengths. Notably, future space-based gravitational wave detectors will offer enhanced sensitivity to the graviton mass \cite{Will:1997bb}.

In a related study, Nishizawa and Arai investigated the modified propagation of gravitational waves in a cosmological context using a parametrized framework, which accounted for a running Planck mass, a modified speed of gravitation, and anisotropic source terms \cite{Nishizawa:2017nef, Arai:2017hxj, Nishizawa:2019rra}. Our results, representing a theory-specific analysis within the Brane-Chern-Simons massive gravity theory, complement their work and provide a foundation for future detailed analysis.

\section{Conclusion}\label{sec:5}

In this work, we introduce a novel extension of massive gravity theory, the Brane-Chern-Simons massive gravity theory. We explore its potential to explain the late-time acceleration of the Universe by deriving the full set of equations of motion for an FLRW background and analyzing self-accelerating background solutions. A key finding is the identification of self-accelerating solutions characterized by an effective cosmological constant, $\tilde{\rho}_{g}$. These solutions exhibit well-behaved behavior without strong coupling issues, offering a promising framework to explain the accelerated expansion of the Universe.

Furthermore, our analysis of tensor perturbations provides insights into the nature of the graviton mass within this theory, leading to the derivation of the dispersion relation for gravitational waves. We also examine the stability of long-wavelength gravitational waves, where the sign of the mass square plays a crucial role. A positive mass square indicates stability, while a negative value suggests the presence of tachyons. However, the development of any instability due to tachyonic behavior is mitigated by the long timescales associated with the Hubble scale.

The calculation of the dispersion relation in Brane-Chern-Simons massive gravity theory offers a means to test gravitational perturbation propagation in FLRW cosmology. Our analysis provides a foundation for future studies, complementing existing work on modified gravitational wave propagation and potentially shedding light on the nature of gravity.

\acknowledgments

This work is based upon research funded by the University of Tabriz, Iran National Science Foundation (INSF), and Iran National Elites Foundation (INEF), under project No. 4014244. 
This work is supported by the National Key R$\&$D Program of China (grant 2023YFE0117200) and the National Natural Science Foundation of China (Nos. 12105013).
The authors thank A. Emir Gumrukcuolu for insightful comments. We also appreciate Nishant Agarwal for sharing notes and computational codes related to tensor perturbations, which greatly facilitated our research.

\paragraph{Note added.} This is also a good position for notes added
after the paper has been written.


\end{document}